\documentclass[fleqn,twoside]{article}
\usepackage{espcrc2}

\usepackage{graphicx}
\usepackage[figuresright]{rotating}

\newcommand{\AmS}{{\protect\the\textfont2
  A\kern-.1667em\lower.5ex\hbox{M}\kern-.125emS}}

\def\neut{{$\nu$}}
\def\neute{{$\nu_e$}}
\def\neutmu{{$\nu_{\mu}$}}

\title{The AMANDA-II Neutrino Telescope}

\author{R.~Wischnewski \address{DESY-Zeuthen, 
Platanenalle 6, D-15735 Zeuthen, Germany}
for the AMANDA-Collaboration
}
       
\begin{document}

\begin{abstract}
The AMANDA-II 
telescope at the South Pole is
constructed of 677 optical modules at 19 strings.
We describe the 
observation of 
atmospheric neutrinos with the first stage 
10-string detector AMANDA-B10, which
establishes AMANDA as a working neutrino telescope.
The expected performance 
for the 
AMANDA-II detector
is discussed.
\vspace{1pc}
\end{abstract}

\maketitle

\section{INTRODUCTION}

To detect the feeble flux of predicted high energy astrophysical 
neutrinos,
high energy Cherenkov neutrino telescopes are build 
as a grid of photomultiplier tubes (PMTs) 
covering a huge geometric volume.
The density of PMTs per effective detector area is about 3 orders 
of magnitude below 
that for \neut-events triggered in the 
SuperKamiokande detector; 
this implies a detailed verification of the 
anticipated performance of these new detectors.

\begin{figure}[h]
\vspace*{-2.0mm} 
\includegraphics[height=9cm]
{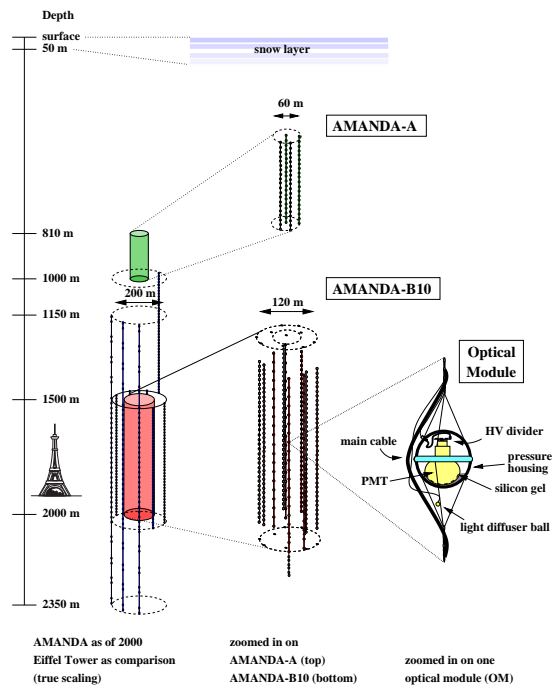} 
\vspace*{-6.0mm} 
\caption{The AMANDA-II detector.}
\vspace*{-.5cm}
\label{fig:detector_layout}
\end{figure}

The AMANDA-II telescope is located at the geographic South Pole
and uses the transparent ice of the 3~km thick ice sheet
\cite{Naturepaper}.
The detector has been 
installed  
between November 1995 and February 2000, 
and consists of 677 PMTs in optical modules (OM) 
on 
19 vertical strings deployed to depths between 1300\,m-2400\,m.
The main instrumented volume ranges from 1500\,m-2000\,m,
it
covers 
a cylinder of 200\,m diameter
(fig.\ref{fig:detector_layout}).

Strings 1-10 form the inner detector of 120\,m  diameter 
called AMANDA-B10, which was commissioned in 1997.
They use passive OMs with electrical analog
signal transmission over 2\,km cable.
Strings 11-19 form the outer cylinder and 
are based on analog fiber transmission
\cite{Icrc01_aii}.

First physics results obtained with AMANDA-B10 
are presented 
at this conference by  \cite{taup_hallgren}.

\section{\neutmu-SEPARATION IN AMANDA-B10}

With no astrophysical neutrino sources 
detected yet,
atmospheric neutrinos are the only source to calibrate the 
sensitivity to 
the favoured \neut-channel: 
upgoing $\mu$'s generated in charged current (CC) 
\neutmu-interactions.

\begin{figure}[h]
\includegraphics[width=6.0cm]
{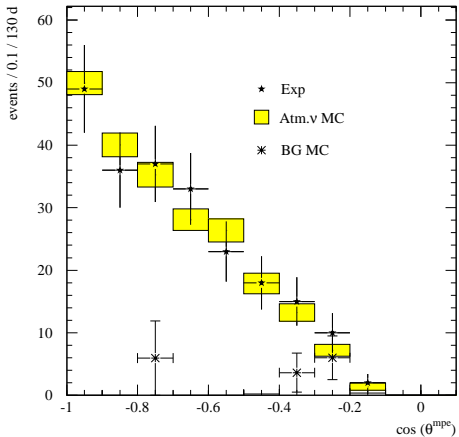} 
\label{fig:cthe_B10}
\vspace*{-9.0mm} 
\caption{Zenith angle distribution for a selected neutrino
sample of 223 events. 
Hatched boxes are MC expectation for atmospheric neutrinos.}
\end{figure}

Muon track reconstruction is based on a 
maximum likelidood fit of measured photon arrival times to 
expected arrival time distributions 
\cite{Icrc01_nuat}\cite{B10nupaper},
and uses 
modeled
optical bulk ice properties 
(light scattering and absorption), 
and local hole ice effects. 
Light emitted by  
muon-induced electromagnetic showers and 
accompanying muons (not in the track fit model)
and the low information density per track 
result in 
downgoing atmospheric muons 
being 
sometimes
misreconstructed as upward tracks.
They form the dominant background to \neutmu's, 
since they trigger about 10$^5$ more abundant.
 
The 1997 data set 
of the AMANDA-B10 detector has 
been analyzed by two independent groups.
They used 
different modifications of track reconstruction
and extended cuts 
for selecting  highest quality upward reconstructed muon tracks
to reject the background at acceptable signal efficiencies. 
From a total sample of 10$^9$ events triggered during 
130 effective livetime days in 1997,
final samples of 223 and 204 events have been extracted,
with  \neutmu\, passing rates of 3-4\% and 
residual background contamination of 10\%
\cite{Icrc01_nuat}\cite{B10nupaper}.
The observed number of events is
consistent with expectation,
within a
$\sim$50\% systematic error 
due to ice property and primary flux uncertainties.
In figure \ref{fig:cthe_B10} the
distribution in cosine of the zenith angle 
after final cuts for the sample of analysis A is shown 
(compared to \neut-simulation expectation, normalized 
to 
experimental data). 
Angular acceptance of AMANDA-B10 is seen to be 
worst
for horizontal directions. 
The angular and energy averaged effective muon area 
for muon energies of  0.1-1\,TeV is 2800\,m$^2$ 
(weighted over the atmospheric \neut-spectrum).
Much higher effective areas are observed for 
other searches (see below).

\section{PERFORMANCE OF AMANDA-II}

The AMANDA-II telescope is in stable operation since 
its commissioning in February, 2000.
Reasonable
agreement is found between data and simulation \cite{Icrc01_aii}.

\begin{figure}[tbp]
\includegraphics[width=6.5cm]
{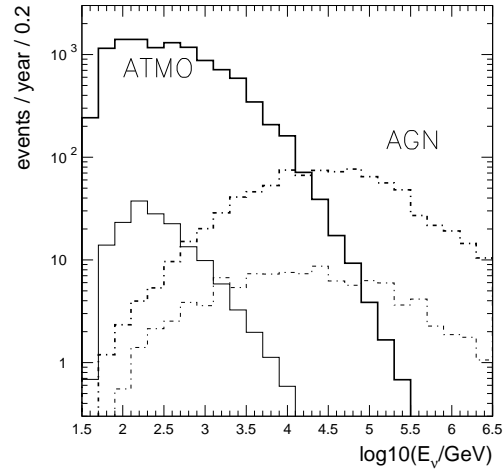}
\vspace*{-9.0mm} 
\caption{Neutrino energy spectrum of events, triggering AMANDA-II. 
Shown are charged current $\nu_{\mu}$ events (thick line) 
and charged current $\nu_e$ events (thin line) 
for atmospheric neutrinos (full line) and an AGN like spectrum (dashed line). 
Event rate is calculated per livetime year.
}
\label{fig:trig_ener}
\end{figure}

For neutrinos from astrophysical sources 
(such as Active Galactic Nuclei, AGN), 
harder spectra than for atmospheric neutrinos are expected.
Fig.\ref{fig:trig_ener} shows the expected spectrum of 
triggered events for atmospheric neutrinos (full line) and for a
spectrum of AGN neutrinos 
(dashed line, equal flux set for $\nu_{\mu}$ and $\nu_{e}$)
with an assumed diffuse flux of
$10^{-6}\cdot E^{-2}$~GeV sr$^{-1}$ s$^{-1}$cm$^{-2}$.
Shown are  CC $\nu_{\mu}$ and CC $\nu_{e}$ reactions.
The total number of triggered upgoing events per livetime year
for the different channels 
are given in table \ref{tb:s_rate},
including neutral current (NC) interactions.

The \neute-channel was recently shown to
yield good sensitivity 
for diffuse high energy \neut-search \cite{taup_domo}\cite{Icrc01_cascades}. 
A first  detection of the 
atmospheric neutrino induced cascades is an interesting challenge
in view of the low event rate, see \cite{Icrc01_cascades}.

\begin{table}[t]
\begin{center}
\begin{tabular}{ | c | c | c |}
\hline
 & Atmospheric   & AGN  \\
\hline
$\nu_{\mu}$ & 11000 (CC), 130 (NC) &   853 (CC) \\
$\nu_{e}$ & 162 (CC), 8.8 (NC) &  103 (CC)   \\
\hline
\end{tabular}
\vspace*{2.0mm} 
\caption{Triggered \neut-events per year in AMANDA-II.
}
\label{tb:s_rate}
\end{center}
\vspace*{-7.0mm} 
\end{table}

\begin{figure}[t]
\includegraphics[width=7.0cm,height=6cm]
{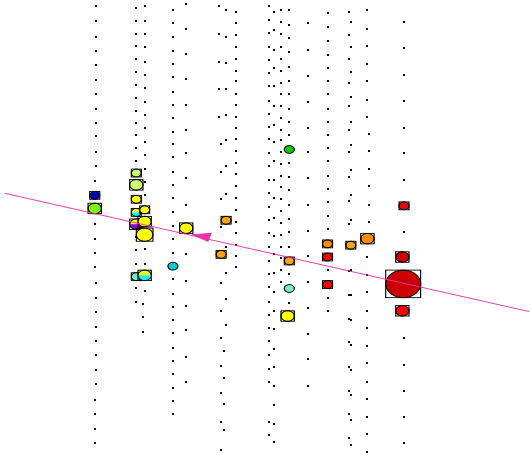} 
\vspace*{-9.0mm} 
\caption{An upward reconstructed AMANDA-II event close to the horizon.
Shown is the central part of the detector, 
colorscale and symbolsize correspond to hit time and amplitude.}
\label{fig:evt_2218020}
\vspace*{-4.0mm} 
\end{figure}

The analysis of AMANDA-II data from 2000 is under way.
With \neutmu-selection cuts tailored 
to the larger detector,
we 
find the angular acceptance for horizontal directions and 
the total signal passing rate much improved, compared to AMANDA-B10.
Fig.\ref{fig:evt_2218020} displays
a good candidate event for an upward moving muon, 
reconstructed close to the horizon (zenith angle = 105$^o$).

\begin{figure}[t]
\includegraphics[width=7.5cm,height=6.5cm]
{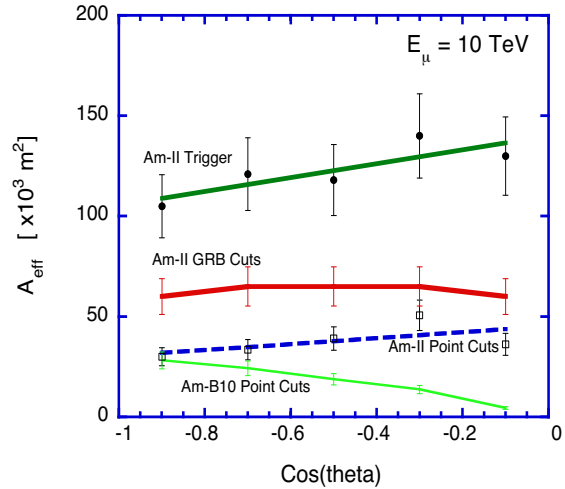}
\vspace*{-11.0mm} 
\caption
{Effective area versus zenith angle for AMANDA-II for muons 
at 10\,TeV,
at trigger level and for GRB and point source analysis.
}
\label{fig:Aeff_aii_costh}
\vspace*{-2.0mm} 
\end{figure}

In figure \ref{fig:Aeff_aii_costh},
the calculated AMANDA-II effective area for 10\,TeV muons 
is shown 
as function of the zenith angle 
at trigger level \cite{Icrc01_phys}. 
Expected effective areas for two high energy  physics analyses 
are also shown: GRB and point source cuts, 
yielding 
areas of 30-60.000\,m$^2$.
The curve shown for the AMANDA-B10 reveals the acceptance improvement  
in horizontal direction.


\section{CONCLUSION}
The AMANDA-II telescope reaches
effective muon detection areas at 10\,TeV   
of up to  50000\,m$^2$,
depending on the physics objective,  
and offers better angular coverage than 
AMANDA-B10.
The latter was proven to work as a high energy neutrino detector
by the isolation of 
a combined sample of 325 atmospheric \neutmu's. 
The rate of atmospheric \neutmu\, events
for the full AMANDA-II detector will be about 3 times higher,
yielding 
800-1000 \neutmu's for year 2000. 
Flux limits obtainable by AMANDA-II for high energy neutrinos 
from diffuse and point sources
will improve by $\sim$10 times within 2 years of livetime
\cite{Icrc01_phys}.

\end{document}